# Evaluation of a Multi-Resolution Dyadic Wavelet Transform Method for usable Speech Detection

Wajdi Ghezaiel, Amel Ben Slimane Rahmouni, and Ezzedine Ben Braiek

*Abstract*—Many applications of speech communication and speaker identification suffer from the problem of co-channel speech. This paper deals with a multi-resolution dyadic wavelet transform method for usable segments of co-channel speech detection that could be processed by a speaker identification system. Evaluation of this method is performed on TIMIT database referring to the Target to Interferer Ratio measure. Co-channel speech is constructed by mixing all possible gender speakers. Results do not show much difference for different mixtures. For the overall mixtures 95.76% of usable speech is correctly detected with false alarms of 29.65%.

## I. INTRODUCTION

Most of speaker recognition systems are developed using clean speech training data. In practical scenario, speech signals can be corrupted by two types of interference, background noise, or another speaker's speech. The performance of speaker recognition systems is known to be adversely affected by the presence of such interferences. Various techniques exist for reduction or elimination of noise distortions in speech signals. However, due to the non-stationary properties of speech, complete removal of speech interferences has been a challenge to the speech processing industry. Speech interference occurs when two or more speakers are speaking simultaneously over the same channel without a significant difference in their overall energy. The resulting speech is commonly termed "co-channel speech".

Historically, the goal of co-channel research has been to be able to extract the speech of one of the talkers from the co-channel speech. This can be achieved by either enhancing the target speech or suppressing the interfering speech [1]. This co-channel situation has presented a challenge to speech researchers for the past 30 years [2] [3] [4]. When the extracted speech is used in a speaker identification system, then one must determine how much and what type of target speech is needed to perform "good" speaker identification, i.e., how much interference is acceptable. Therefore, determining the effect of speaker interference on speaker identification would be of considerable interest. Various methods have been proposed in the literature to improve the robustness of a speaker recognizer to overcome this co-channel problem [5] [6] [7] [8]. Usable speech extraction is a novel concept of processing degraded speech data. When the energies of the target and interferer speech are approximately equal, certain portions of speech still exist in co-channel speech in which the energy of one speaker is greater than the energy of the other speaker. These portions are termed "usable" while the other portions are termed "unusable". The use of only 'usable' portions of speech has been shown improve the performance of speaker identification systems [5] [6]. Yantorno [7] performed a study on co-channel speech and concluded that the Target-to-Interferer Ratio (TIR) was a good measure to quantify usability for speaker identification. Several methods based on usable speech measures which have referred to the TIR measure have been developed and studied under co-channel conditions [9] [10] [11] [12] [13]. These methods show that the speaker identification system could achieve approximately 80% of correct identification when the overall TIR is 20 dB.This paper presents evaluation of a novel method for usable speech detection based on a multi-resolution dyadic wavelet transform MRDWT [14]. In order to detect usable speech, the proposed method MRDWT looks for hidden periodicity feature. If a co-channel speech is usable, periodicity feature is not disturbed by interferer speech. A multi-resolution analysis could easily extract periodicity feature from all lower frequency sub-band of co-channel speech. In fact, wavelet transform has shown good capability to detect the abrupt changes in the amplitude level of the speech signal, e.g., for pitch detection, [15] [16], or for voiced/unvoiced detection [15]. Periodicity features are localized by applying discrete wavelet transform (DWT) iteratively to achieve the dominant frequency band of pitch. Evaluation of this method is performed on TIMIT database referring to the TIR measure. Co-channel speech is constructed by mixing all possible gender speakers. Discussion of the merits and limitations of the proposed method are provided basing on evaluation results.The rest of this paper is organized as follows. In Section 2, we describe the multi-resolution dyadic wavelet transform method for usable speech detection. In Section 3, we describe the proposed evaluation system. In Section 4, we report the

Wajdi Ghezaiel and Ezzedine Ben Braiek are with the CEREP- ESSTT, Tunis university, Tunis, Tunisia (e-mail:wajdi.ghezaiel@gmail.com Ezzedine.Benbraiek@esstt.rnu.tn).
Amel Ben Slimane Rahmouni, is with ENSI, Mannouba University, Mannouba, Tunisia. (e-mail: Amel.benslimane@ensi.rnu.tn).

experimental results and discussions. Finally, we present the conclusions.

## II. MULTI-RESOLUTION DYADIC WAVELET TRANSFORM METHOD FOR USABLE SPEECH DETECTION

It has been shown that usable speech cannot be determined from unvoiced frames [6]. Hence, voiced frames are detected by applying a DWT on the co-channel speech. These frames are characterized by the energy of approximation coefficients witch occupies more than 90% of overall energy [17]. Usable frames are then characterized by periodicity features. These features should be located in low-frequency band that includes the pitch frequency. Multi-resolution analysis based on DWT [18] [19] is applied iteratively in order to determine the suitable band for periodicity detection In this band periodicity features are not much disturbed by interferer speech in case of usable segments. In case of unusable frames it is not possible to detect periodicity in all lower sub-bands until the pitch band of a male speaker. Pitch band of a male speaker is lower than a female speaker pitch band so that it is imperative to filter until this band. At each scale, autocorrelation is applied to the approximation coefficients in order to detect periodicity. Three maxima are determined from the autocorrelation signal with a peak-picking algorithm using threshold on local maximum amplitude. A difference of autocorrelation lag between the first and second maxima and the autocorrelation lag between the second and third maxima is determined. This difference value notices the presence of pitch information. If this difference is less than the preset threshold periodicity is detected. In our experiment speech signal is sampled at 16000 Hz. We have sub-sampled to 8000 Hz so that the bandwidth of the input speech signal is 4000 Hz. As the pitch frequency bandwidth of a male speaker is approximately limited to 250 Hz for a male speaker, the order of DWT decomposition can be computed up to equate 4. Analysis is performed on frame by frame basis. A frame length is of 64 ms. The Haar wavelet is used in multi-resolution analysis. The following figures present analysis of co-channel frame speech. Figure 1 corresponds to a usable frame for male-male co-channel speech. Periodicity is detected at scale 1. Figure 2 shows that periodicity is not detected at scale 1 but it is detected at scale 2. Figure 3 corresponds to a usable frame for male-male co-channel speech. In this case, Periodicity is detected only at scale 3.

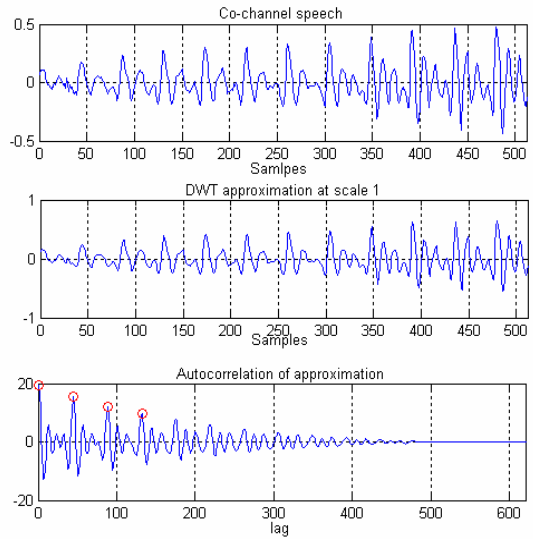

Fig. 1 Analysis of a usable speech frame for male-male co-channel speech up to scale 1

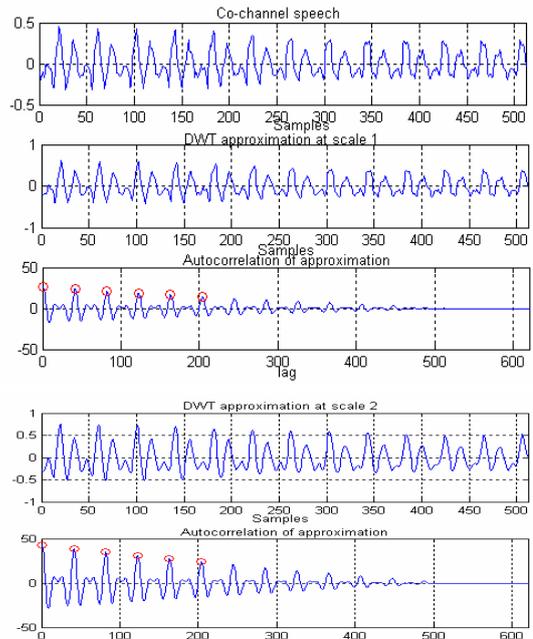

Fig. 2 Analysis of a usable speech frame for female-female co-channel speech up to scale 2.

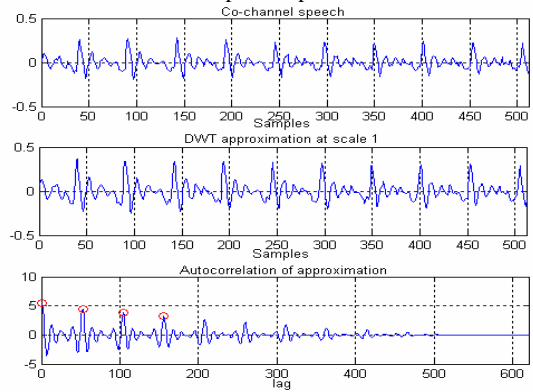

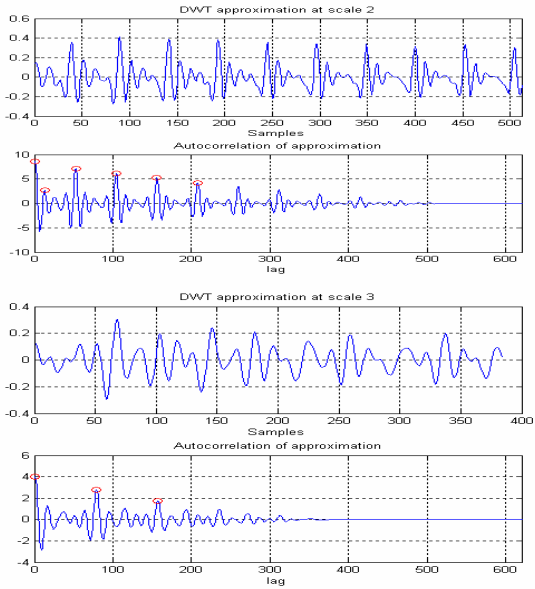

Fig. 3 Analysis of a usable speech frame for male-male co-channel speech up to scale 3.

Figure 4 corresponds to a usable frame for male-male co-channel speech. In this case, Periodicity is detected only at scale 4. Figure 5 presents unusable frame speech. This frame is classified us unusable because periodicity was not detected at all scales.

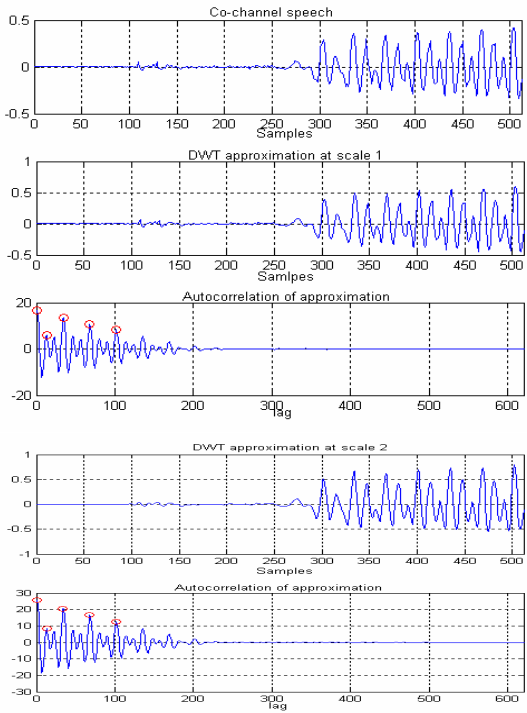

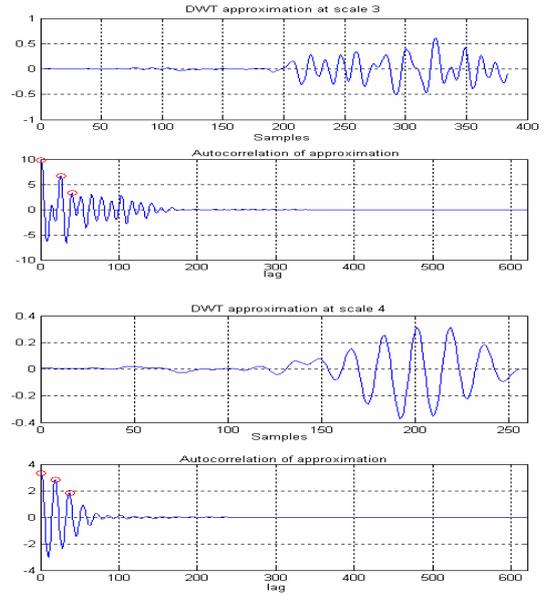

Fig. 4 Analysis of a usable speech frame for male-male co-channel speech up to scale 4

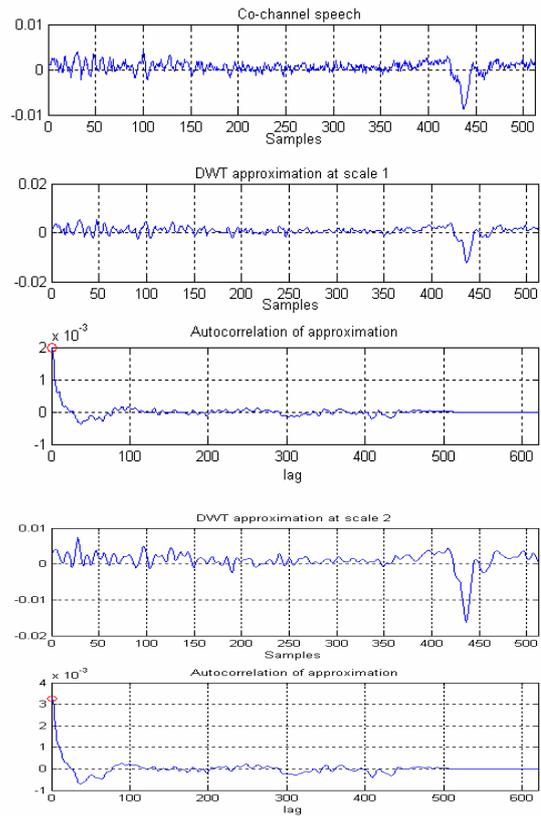

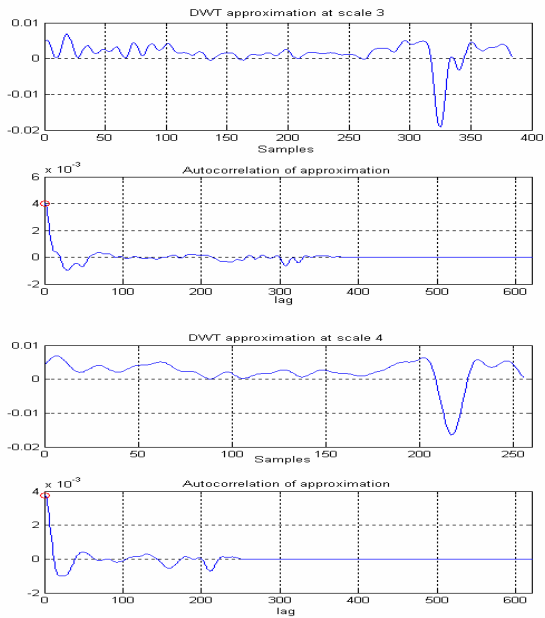

Fig. 5 Analysis of a unusable speech frame up to scale 4.

## III. EXPERIMENTS AND RESULTS

Co-channel speech is created by combining speech files for each of 26 speakers (13 male, 13 female) from the TIMIT database. Co-channel speech is realized by adding speech signals corresponding to two different talkers with an overall TIR of 0 dB. Three different sets of co-channel speech are considered: male-male, female-female, and male-female.

The Target to Interferer Ratio TIR measure is used to label voiced frames as usable or unusable. For usability decision,

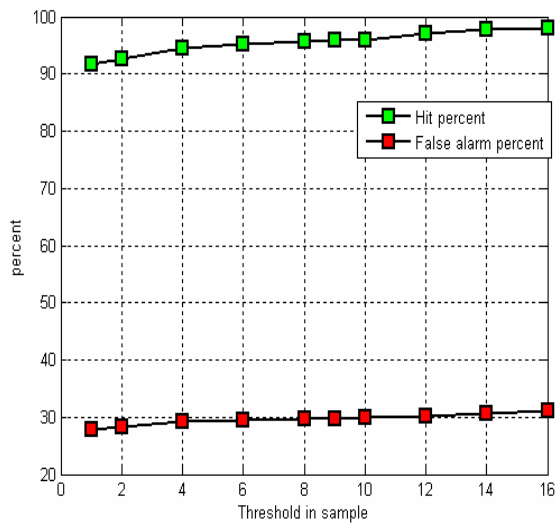

Fig. 6 MRDWT results against threshold values

frames that have above 20 dB TIR are considered as usable. Evaluation is based on hits and false alarms percentages. Decision provided by the MRDWT method is said to have a hit if as well as TIR a frame is labeled as usable. This decision is said to have a false alarm if a frame is classified as usable by the MRDWT method, but labeled as unusable referring to TIR.

Choice of the appropriate threshold preset for usability decision is based on evaluation results. Figure 6 shows the percentage of hits and false alarms versus the threshold value. It is considerable the trade-offs associated with selecting the threshold. For example, selecting a low threshold value will ensure fewer false alarms; however, the trade-off is that less of the co-channel speech will be flagged as usable speech by the proposed method. By examining Figure 6 it is clear that raising the threshold value will flag more speech as usable, and consequently, more false alarms will occur. It can also be observed from Figure 6 that although there is a linear increase in percentages of correct detection and false detection as the threshold is increased, there is a significant increase in false alarm rate when threshold was changed from 8 samples to 16 samples with only a slight increase in correct detection. Therefore the optimum threshold value of 8 samples is chosen.

Figure 7 shows usable speech detection by MRDWT method, for a female-female speech. A few labeled unusable frames are classified as usable and generating false alarms. A closer look at these false alarms reveals that most of them lie in a transition frame from voiced to unvoiced/silence.

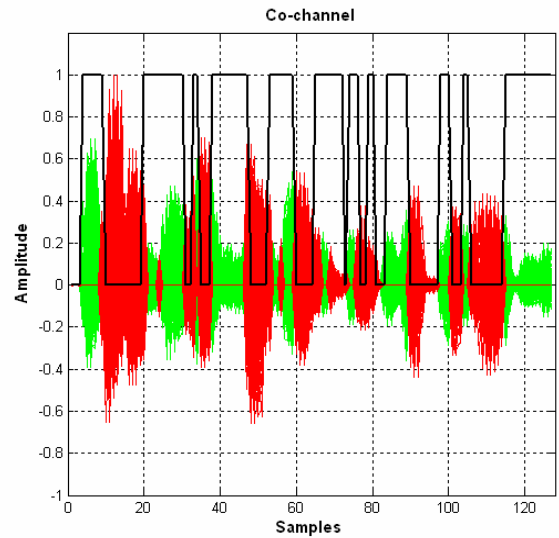

Fig. 7 Detection of usable speech: usable speech (green), unusable speech (red), detected usable speech (+1 rectangles)

As shown in Table 1, the MRDWT does not show much difference in percent of hit detection for male speech and female speech. On an average the proposed method for usable speech detection system detects around 95.76% of usable speech with false alarms around 29.65%.

## IV. CONCLUSION

In this paper, we have presented a new method for usable speech detection based on multi-resolution analysis using a dyadic wavelet transform MRDWT. The goal of the MRDWT method is to extract the maximum amount of usable speech segments with a minimum false alarm rate. The MRDWT method gives good hits percentage, but false alarms have to be improved.

TABLE I
RESULTS OF MRDWT METHOD OF USABLE SPEECH DETECTION

| Co-channel Speech | % Hit | % False alarm |
| --- | --- | --- |
| Female-Female | 93.02 | 32.37 |
| Male-Male | 98.46 | 28.93 |
| Male-Female | 95.80 | 27.66 |
| Average | 95.76 | 29.65 |